%% file: arxiv.tex
\documentclass[journal]{IEEEtran}
\usepackage[OT1]{fontenc}
\usepackage{amsmath,amssymb,amsthm,bm,graphicx,booktabs,array,cite,url}
\usepackage{xcolor}
\usepackage[colorlinks=true,citecolor=blue,linkcolor=blue,urlcolor=blue]{hyperref}
\let\OriginalEqref\eqref
\renewcommand{\eqref}[1]{\textcolor{blue}{\OriginalEqref{#1}}}

\newcommand{\CN}{\mathcal{CN}}
\newcommand{\E}{\mathbb E}

\newcommand{\I}{\mathbf I}
\newcommand{\HH}{\mathbf H}
\newcommand{\K}{\mathbf K}
\newcommand{\Sig}{\mathbf\Sigma}
\DeclareMathOperator{\tr}{tr}
\DeclareMathOperator{\diag}{diag}

\begin{document}
\ifCLASSOPTIONonecolumn
  \setlength{\emergencystretch}{1em}
\else
  \setlength{\abovedisplayskip}{4pt plus 2pt minus 1pt}
  \setlength{\belowdisplayskip}{4pt plus 2pt minus 1pt}
  \setlength{\abovedisplayshortskip}{0pt plus 2pt}
  \setlength{\belowdisplayshortskip}{3pt plus 2pt minus 1pt}
\fi
\title{Exact Error Analysis of Noise-Domain OTFS with Paired Covariance Signaling}
\author{Mahmoud Aldababsa and Khalid Yahya}
\maketitle
\begin{abstract}
Noise-domain orthogonal time frequency space (ND-OTFS) encodes bits
through opposite correlations in Gaussian delay--Doppler pairs of
equal average energy. This letter derives the exact conditional bit
error probability (BEP) of maximum-likelihood detection using
generalized covariance eigenvalues and characteristic-function
inversion. For scalar channels, we obtain an incomplete-beta
expression and its finite-sum Rayleigh average. Partial correlation
yields an error floor for a fixed observation count. At unit
correlation, the conditional BEP decays with an signal-to-noise ratio (SNR) exponent equal
to the number of pair observations, and common Rayleigh fading yields
diversity order one. Simulations validate the analysis and quantify
pair-placement effects.
\end{abstract}

\section{Introduction}
\IEEEPARstart{O}{rthogonal} time frequency space (OTFS) modulation represents information and channel structure on a delay--Doppler (DD) grid. Conventional OTFS assigns constellation symbols to DD resources. The input--output relation depends on pulse shaping, delay support, and Doppler sampling, and practical waveforms can introduce interference beyond an ideal two-dimensional circular-shift model \cite{raviteja}.

Noise modulation encodes information in the statistics of a random waveform. The transmitter selects a statistical state for each message, and the receiver infers that state from the received samples. Variance and correlation are examples of information-bearing statistics. Correlation allows states to have the same marginal sample powers and average transmit energy.
Thermal-noise communication \cite{thermal} and NoiseMod \cite{basar} establish foundations for statistical signaling. Differential binary noise modulation encodes bits in polarity transitions of a reused waveform \cite{dbn}.

Noise-domain OTFS (ND-OTFS) uses a message-dependent covariance for the random DD frame \cite{ndotfs}. This framework includes variance-only, correlation-domain, and separable DD covariance alphabets. In the paired construction studied here, a binary message controls the sign of the correlation within each DD pair. Conditioned on the bit, Gaussian pairs are drawn independently across pair indices and frames.

Exact error analysis is needed to determine how covariance geometry, DD channel mixing, and the number of observations affect detection. Finite-sample overlap between full-rank Gaussian states can persist at high signal-to-noise ratio (SNR), and repeating a state consumes resources that could carry independent messages. The effective matrix channel governs the received covariance, so a scalar SNR alone generally does not characterize multipath detection.

\textcolor{black}{Prior ND-OTFS work \cite{ndotfs} introduces equal-energy covariance signaling and describes Gaussian maximum-\allowbreak likelihood (ML) detection. Complex Gaussian quadratic-form methods have also been used for exact conditional error analysis of differential binary noise modulation \cite{dbndiversity}. At $r=1$ under a common scalar channel, a single bit decision in our model has the same Gaussian covariance hypotheses and correlation statistic as single-branch differential binary noise modulation \cite{dbn,dbndiversity}, after marginalizing the Gaussian waveform. Here the independent pairs occupy DD resources, whereas differential signaling reuses a waveform across successive intervals. This letter makes four contributions.}

\textcolor{black}{\textit{First,} we formulate equal-energy binary ND-OTFS using opposite within-pair correlations. The two states have identical average useful-frame energy and identical per-bin powers. The correlation magnitude controls the separation of the states, and a permutation specifies where the pairs lie on the DD grid.}

\textcolor{black}{\textit{Second,} we derive the exact conditional BEP of the binary ML detector for a known effective DD channel. The generalized eigenvalues of the two received covariances determine the decision threshold and the signed-gamma weights, which give the two error tails through characteristic-function inversion. The same calculation gives pairwise terms for union bounds on larger covariance alphabets.}

\textcolor{black}{\textit{Third,} for a scalar channel, we derive an incomplete-beta BEP and its finite-sum average under common Rayleigh fading. We quantify the positive error floor at fixed observation count for partial correlation. At unit correlation, the conditional SNR exponent equals the number of pair observations, and the common-Rayleigh diversity order is one, with a logarithmic correction for a single pair observation.}

\textcolor{black}{\textit{Fourth,} we check the analysis by Monte Carlo and by sampled-waveform simulation with explicit cyclic-prefix insertion and removal. At equal payload and useful-frame energy, the exact ordering of delay and Doppler pairing reverses with SNR for one fixed fractional-Doppler channel. The tested 64-channel ensemble gives no clear evidence of an average pairing advantage.}

\section{Paired Signaling and Channel Model}\label{sec:model}
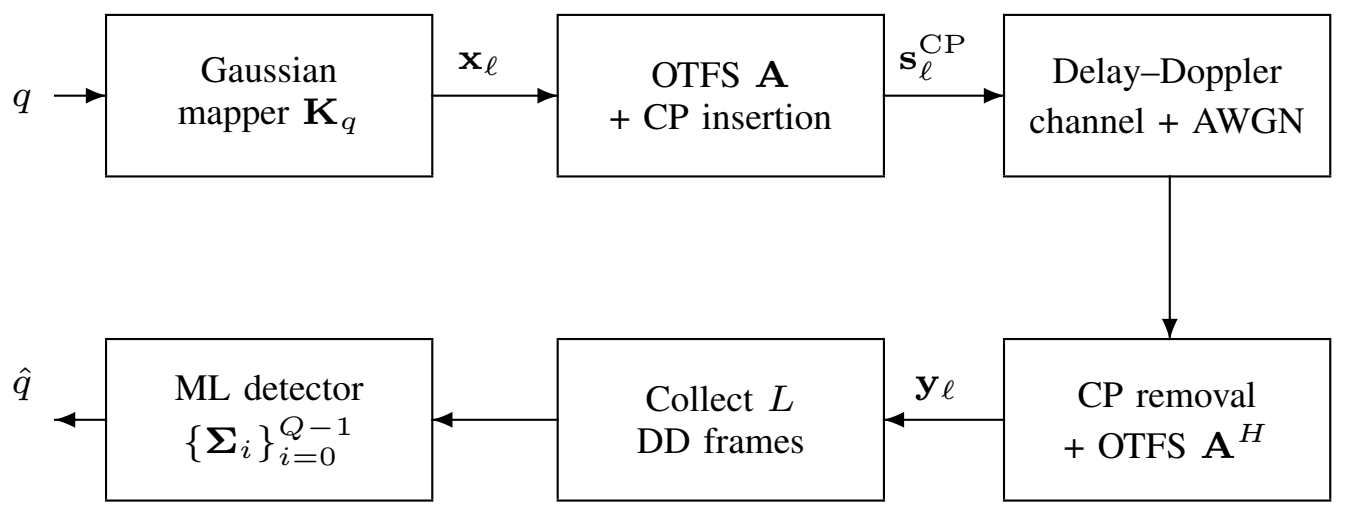
\begin{figure}[!t]
\centering
\input{figures/transceiver.tex}
\caption{\textcolor{black}{Covariance-signaling OTFS transceiver for a $Q$-state alphabet.}}
\label{fig:system}
\end{figure}

\subsection{Statistical DD Signaling}
An OTFS frame contains $M$ delay bins and $N$ Doppler bins, yielding
$N_{\mathrm{DD}}=MN$ DD resources. A message
$q\in\{0,\ldots,Q-1\}$ selects a normalized covariance matrix
$\K_q$ from a fixed, channel-independent alphabet of size $Q$.
The same message is conveyed over $L$ frames. Conditioned on $q$
and the effective channel $\HH$, the transmit vectors
$\mathbf x_\ell\in\mathbb C^{N_{\mathrm{DD}}}$ are drawn independently
according to
\begin{equation}\label{eq:tx}
\mathbf x_\ell\mid q,\HH
\sim\CN(\mathbf 0,P\K_q),\qquad \ell=1,\ldots,L,
\end{equation}
where $\CN$ denotes a proper complex Gaussian distribution,
$\K_q=\K_q^H\succeq 0$, and the superscript $H$ denotes conjugate
transpose. \textcolor{black}{Fig.~\ref{fig:system} shows the general $Q$-state model. Section~\ref{sec:paired} specializes it to binary paired signaling.}

The normalization $\tr(\K_q)=N_{\mathrm{DD}}$ gives every message
the same average useful-frame energy $PN_{\mathrm{DD}}$.
The additional constraint $\diag(\K_q)=\mathbf 1_{N_{\mathrm{DD}}}$
sets the average energy of each DD component to $P$. Consequently,
the covariance states can differ in their off-diagonal entries
while sharing the same marginal powers. For $Q=2^m$ with fixed-length
$m$-bit labels, one message occupies $LN_{\mathrm{DD}}$ DD resources.
The payload is therefore $m/(LN_{\mathrm{DD}})$ bits per DD resource,
excluding overhead.

For frame $\ell$, the receiver observes the DD-domain vector
$\mathbf y_\ell\in\mathbb C^{N_{\mathrm{DD}}}$ given by
\begin{equation}\label{eq:rx}
\mathbf y_\ell=\HH\mathbf x_\ell+\mathbf n_\ell,
\end{equation}
where $\HH\in\mathbb C^{N_{\mathrm{DD}}\times N_{\mathrm{DD}}}$
is known at the receiver and is assumed constant over the $L$ frames.
The noise vectors $\mathbf n_\ell$ are independent across frames
and independent of the transmitted vectors. Each satisfies
$\mathbf n_\ell\sim\CN(\mathbf 0,N_0\I_{N_{\mathrm{DD}}})$,
where $N_0>0$ is the noise variance per complex DD sample.

Conditioned on $q$ and $\HH$, the signal and noise have zero mean,
so $\mathbb E[\mathbf y_\ell\mid q,\HH]=\mathbf 0$. Their independence
eliminates the cross terms in the received covariance, giving
\begin{align}
\Sig_q
&=\mathbb E[\mathbf y_\ell\mathbf y_\ell^H\mid q,\HH]\nonumber\\
&=\HH\,\mathbb E[\mathbf x_\ell\mathbf x_\ell^H\mid q,\HH]\HH^H
  +\mathbb E[\mathbf n_\ell\mathbf n_\ell^H]\nonumber\\
&=P\HH\K_q\HH^H+N_0\I_{N_{\mathrm{DD}}}.\label{eq:sigma}
\end{align}
Thus, $\mathbf y_\ell\mid q,\HH\sim\CN(\mathbf 0,\Sig_q)$,
independently across frames. The effective channel maps the
transmit covariance $\K_q$ into the received signal covariance
$P\HH\K_q\HH^H$, which determines how distinguishable the covariance
states are at the receiver. Since
$\Sig_q\succeq N_0\I_{N_{\mathrm{DD}}}\succ 0$, its inverse and log
determinant are well defined even when $\K_q$ or $\HH$ is rank deficient.

If the raw DD-domain noise has known covariance
$N_0\mathbf R_n$ with $\mathbf R_n\succ 0$, the receiver applies
$\mathbf R_n^{-1/2}$ to the raw observations. In this case,
$\HH=\mathbf R_n^{-1/2}\mathbf H_{\mathrm{raw}}$ in
\eqref{eq:rx}, and the prewhitened noise has covariance
$N_0\I_{N_{\mathrm{DD}}}$. We define $\rho=P/N_0$ as the nominal
transmit SNR per DD resource.

\subsection{Effective OTFS Channel Model}
We consider rectangular-pulse OTFS with a reduced cyclic prefix (CP),
using one CP per frame \cite{raviteja_rcp}. With the DD samples stacked column-wise in
$\mathbf x_\ell$, the useful time-domain samples are
$\mathbf s_\ell=\mathbf A\mathbf x_\ell$, where
$\mathbf A=\mathbf F_N^H\otimes\I_M$ and $\mathbf F_N$ is the unitary
$N$-point discrete Fourier transform matrix. Since $\mathbf A$ is
unitary, the receiver applies $\mathbf A^H$ after CP removal to
recover the DD-domain observation.

Let $S$ denote the number of paths, and let $h_p$, $d_p$, and $f_p$
denote the complex gain at the useful-frame time origin, the integer
sample delay, and the Doppler frequency in hertz of path $p$,
respectively. With sampling interval $T_s$, the normalized Doppler is
$\nu_p=f_pN_{\mathrm{DD}}T_s$, which may be fractional. When the CP
length is at least $\max_p d_p$, the useful receive samples after CP
removal are
\begin{align}
r_\ell[n]
&=\sum_{p=1}^{S}h_p e^{j2\pi\nu_p n/N_{\mathrm{DD}}}
s_\ell[(n-d_p)\bmod N_{\mathrm{DD}}]+w_\ell[n],
\nonumber\\[-1pt]
&\hspace{4em} n=0,\ldots,N_{\mathrm{DD}}-1,
\end{align}
where $w_\ell[n]$ is the time-domain noise. Define the one-sample
circular delay by
$[\mathbf\Pi\mathbf s]_n=s[(n-1)\bmod N_{\mathrm{DD}}]$
and the diagonal Doppler matrix by
$[\mathbf D(\nu)]_{n,n}=e^{j2\pi\nu n/N_{\mathrm{DD}}}$.
Stacking the receive samples in $\mathbf r_\ell$ and applying
$\mathbf y_\ell=\mathbf A^H\mathbf r_\ell$ yields the effective
DD-domain channel
\begin{equation}\label{eq:channel}
\HH=\mathbf A^H
\left(\sum_{p=1}^{S}h_p\mathbf D(\nu_p)\mathbf\Pi^{d_p}\right)
\mathbf A.
\end{equation}
For white time-domain noise with covariance $N_0\I_{N_{\mathrm{DD}}}$,
the DD-domain noise $\mathbf n_\ell=\mathbf A^H\mathbf w_\ell$ has
the same covariance because $\mathbf A^H\mathbf A=\I_{N_{\mathrm{DD}}}$.

Path phases may evolve between frames under fractional Doppler.
For $L>1$, the analysis assumes that the effective channel $\HH$
is fixed over the observation block. All fractional-Doppler
numerical tests use $L=1$.

\section{Optimal Detection and Exact Conditional Error}
\label{sec:conditional}

Let $\mathbf Y=[\mathbf y_1,\ldots,\mathbf y_L]$ collect the $L$
received frames. Conditioned on $q$ and $\HH$, the columns are
independent and each has distribution $\CN(\mathbf 0,\Sig_q)$.
Their joint likelihood is
\begin{equation}
p(\mathbf Y\mid q,\HH)
=
\frac{\exp\!\left(-\sum_{\ell=1}^L
\mathbf y_\ell^H\Sig_q^{-1}\mathbf y_\ell\right)}
{\pi^{LN_{\mathrm{DD}}}(\det\Sig_q)^L}.
\end{equation}
For equiprobable messages, the ML rule
minimizes the negative log-likelihood after dropping terms
independent of $q$:
\begin{equation}
\widehat q=\arg\min_q
\left\{L\log\det\Sig_q+
\sum_{\ell=1}^L\mathbf y_\ell^H\Sig_q^{-1}\mathbf y_\ell\right\}.
\end{equation}
A Cholesky factorization gives the log determinant and allows
evaluation of the quadratic terms by triangular solves.

For binary signaling, the log-likelihood ratio of state~1 to
state~0 is $T-b$, where
\begin{align}
T&=\sum_{\ell=1}^L\mathbf y_\ell^H
(\Sig_0^{-1}-\Sig_1^{-1})\mathbf y_\ell,\\
b&=L\log\frac{\det\Sig_1}{\det\Sig_0}.
\end{align}
The receiver selects state~1 if $T>b$ and state~0 otherwise.

Let
$\Sig_0^{-1/2}\Sig_1\Sig_0^{-1/2}
=\mathbf U\mathbf\Lambda\mathbf U^H$,
where $\mathbf U$ is unitary and
$\mathbf\Lambda=\diag(\lambda_1,\ldots,\lambda_{N_{\mathrm{DD}}})$
contains the positive generalized eigenvalues of $(\Sig_1,\Sig_0)$.
The transformed vector
$\mathbf z_\ell=\mathbf U^H\Sig_0^{-1/2}\mathbf y_\ell$
has covariance $\I_{N_{\mathrm{DD}}}$ under state~0 and
$\mathbf\Lambda$ under state~1. Substituting
$\mathbf y_\ell=\Sig_0^{1/2}\mathbf U\mathbf z_\ell$ gives
\begin{equation}
T=\sum_{\ell=1}^L
\mathbf z_\ell^H(\I_{N_{\mathrm{DD}}}-\mathbf\Lambda^{-1})
\mathbf z_\ell.
\end{equation}
Under state~0, $|z_{\ell,i}|^2$ is unit-mean exponential.
Under state~1, $|z_{\ell,i}|^2/\lambda_i$ has the same
distribution. These normalized terms are independent across
frames and coordinates. Summing over the $L$ frames gives
mutually independent gamma variables $G_{q,i}$ of shape $L$
and unit scale under each state $q$. The conditional
distributions of $T$ are those of
\begin{align}
W_0&=\sum_{i=1}^{N_{\mathrm{DD}}}
(1-\lambda_i^{-1})G_{0,i},\\
W_1&=\sum_{i=1}^{N_{\mathrm{DD}}}
(\lambda_i-1)G_{1,i},
\end{align}
under states~0 and~1, respectively.

Let $F_W(z)=\Pr(W\le z)$ denote the cumulative distribution
function (CDF) of $W$. Taking determinants in the whitening
relation gives $\det(\Sig_1)/\det(\Sig_0)=\prod_i\lambda_i$.
The two conditional error events are $W_0>b$ and $W_1\le b$.
Averaging their probabilities yields
\begin{align}
b&=L\sum_{i=1}^{N_{\mathrm{DD}}}\log\lambda_i,\\
P_b(\HH)&=\tfrac12\bigl[1-F_{W_0}(b)+F_{W_1}(b)\bigr].
\label{eq:matrix}
\end{align}

When $\HH(\K_1-\K_0)\HH^H=\mathbf 0$, the received covariances
coincide and the equal-prior BEP is
$1/2$. Otherwise, at least one $\lambda_i$ differs from unity,
so both weighted sums have continuous CDFs. For either sum
$W=\sum_i w_iG_i$, independence gives the characteristic
function below. The CDF follows from the Gil--Pelaez inversion
formula \cite{gil}:
\begin{align}
\phi_W(t)&=\prod_{i:w_i\ne0}(1-jtw_i)^{-L},\\
F_W(z)&=\frac12-\frac1\pi\int_0^\infty
\frac{\operatorname{Im}\{e^{-jtz}\phi_W(t)\}}{t}\,dt.
\label{eq:inv}
\end{align}
The integrand has the limit $L\sum_i w_i-z$ as $t\to0$.
Equations~\eqref{eq:matrix}--\eqref{eq:inv} accommodate signed
and repeated weights and account for DD mixing through both
received covariances.

For random $\HH$ known to the receiver, the average binary BEP
is $\overline P_b=\E_{\HH}[P_b(\HH)]$. For $Q>2$ with distinct
received covariances, the probability that the metric for
state $v$ beats that for the transmitted state $q$ is the
tail $1-F_{W_0}(b)$ evaluated with $(\Sig_q,\Sig_v)$ in place of
$(\Sig_0,\Sig_1)$. Summing over $v\ne q$ and averaging over the
equiprobable transmitted states gives a symbol error probability
union bound. For $Q=2^m$ with fixed-length $m$-bit labels,
weighting each pairwise term by the Hamming distance between the
two labels divided by $m$ gives a BEP union bound.

\section{Equal-Energy Paired Correlation Signaling}\label{sec:paired}
\label{sec:scalar}

\subsection{Covariance Alphabet and DD Pairing}

For $Q=2$ equiprobable states and even $N_{\mathrm{DD}}=2J$,
let $J$ denote the number of disjoint DD pairs. Define
\begin{equation}\label{eq:codebook}
\K_q=\mathbf V(\I_J\otimes\mathbf B_q)\mathbf V^H,\qquad
\mathbf B_q=
\begin{bmatrix}
1 & (-1)^q r\\
(-1)^q r & 1
\end{bmatrix},
\end{equation}
where $0\le r\le1$ is the correlation magnitude and $\mathbf V$
is a permutation matrix that places the pairs on the DD grid.
Each pair block has eigenvalues $1\pm r$. Both states have unit
diagonal and trace $N_{\mathrm{DD}}$, giving the same average
useful-frame energy $N_{\mathrm{DD}}P$.

For each frame $\ell$ and pair index $j$, draw
$u_{\ell,j},v_{\ell,j}\sim\CN(0,1)$ independently. Before
applying $\mathbf V$, generate the pair as
\begin{equation}
\begin{bmatrix}
x_{\ell,j,1}\\ x_{\ell,j,2}
\end{bmatrix}
=
\sqrt P
\begin{bmatrix}
u_{\ell,j}\\
(-1)^q r u_{\ell,j}+\sqrt{1-r^2}\,v_{\ell,j}
\end{bmatrix}.
\end{equation}
Each entry has variance $P$, and their cross-covariance is
$P(-1)^q r$. Independent draws across pair indices give the
block-diagonal covariance before permutation. At $r=1$, the
second entry is a signed copy of the first.

\subsection{Scalar-Channel Sufficient Statistic}

For $\HH=h\I_{N_{\mathrm{DD}}}$, define
$\gamma=\rho|h|^2$, $\eta=r\gamma/(1+\gamma)$, and
$a=N_0(1+\gamma)$. After undoing $\mathbf V$, each received
pair $(y_1,y_2)$ has marginal variance $a$ and cross-covariance
$(-1)^q a\eta$. The modes $z_\pm=(y_1\pm y_2)/\sqrt{2}$
are independent proper complex Gaussian variables with variances
$a[1\pm(-1)^q\eta]$. Let $E_\pm$ be the sum of $|z_\pm|^2$
over the $K=JL$ pair observations. The two states have equal
covariance determinants. Subtracting their quadratic metrics gives
\begin{equation}
\log\frac{p(\mathbf Y\mid 1,\HH)}{p(\mathbf Y\mid 0,\HH)}
=\frac{2\eta}{a(1-\eta^2)}(E_--E_+).
\end{equation}
For $\eta>0$, the receiver selects state~1 if $E_->E_+$.
Since $|z_+|^2-|z_-|^2=2\operatorname{Re}(y_1y_2^*)$, the
decision depends on the sign of the summed within-pair correlation.

Under state~0, $U=E_+/[a(1+\eta)]$ and
$V=E_-/[a(1-\eta)]$ are independent gamma variables of shape
$K$ and unit scale. The conditional error event is equivalent
to $U/(U+V)<(1-\eta)/2$. To obtain the distribution of this
ratio, set $t=U/(U+V)$ and $s=U+V$. The inverse transformation
$(U,V)=(ts,(1-t)s)$ has Jacobian $s$. Integrating the joint
density over $s$ yields the beta density
$t^{K-1}(1-t)^{K-1}/B(K,K)$ on $0<t<1$, where $B$ is the
beta function. Under state~1, the energy distributions are
exchanged, giving the same conditional error probability.
Therefore,
\begin{align}
P_b(\gamma)&=I_{x(\gamma)}(K,K),\qquad
x(\gamma)=\frac{1+(1-r)\gamma}{2(1+\gamma)},
\label{eq:beta}\\
I_x(K,K)&=\sum_{n=K}^{2K-1}
\binom{2K-1}{n}x^n(1-x)^{2K-1-n},
\end{align}
where $I_x$ is the regularized incomplete beta function
\cite{dlmfbeta}. If $r\gamma=0$, the two states are
indistinguishable and \eqref{eq:beta} gives $P_b=1/2$.

For $0<r<1$, the scalar BEP approaches the positive floor
$I_{(1-r)/2}(K,K)$ as $\gamma\to\infty$. At $r=1$,
$x(\gamma)=1/[2(1+\gamma)]$ and
$I_x(K,K)\sim x^K/[KB(K,K)]$ as $x\downarrow0$. Consequently,
\begin{equation}\label{eq:asym}
P_b(\gamma)\sim\frac{\gamma^{-K}}{2^K K B(K,K)}.
\end{equation}

For fixed invertible $\HH$ and $r<1$, applying
$\HH^{-1}/\sqrt P$ to each received vector yields the
conditional covariance
$\K_q+\rho^{-1}\HH^{-1}(\HH^{-1})^H$, which converges to
$\K_q$ as $\rho\to\infty$. The corresponding joint Gaussian
densities converge in $L^1$ for fixed $L$, so the optimal binary
BEP converges to the same floor as in the scalar channel.
With singular $\HH$, the two received covariances can coincide
even at $r=1$, as for one pair with $\HH=\diag(1,0)$.

\section{Exact Scalar Rayleigh-Averaged Error}
\label{sec:rayleigh}

Let the same scalar coefficient $h\sim\CN(0,\beta)$ apply to all
$K$ pair observations. Then $\gamma=\rho|h|^2$ is exponential
with mean $\bar\gamma=\rho\beta$, and the averaged BEP is
$\bar P_b=\E_\gamma[I_{x(\gamma)}(K,K)]$. Write
$x(g)=c+d/(1+g)$, where $c=(1-r)/2$ and $d=r/2$.
For $m\ge1$, the relevant moments are
\begin{equation}
M_m=\E[(1+\gamma)^{-m}]
=\frac{e^{1/\bar\gamma}}{\bar\gamma}
\mathsf E_m(1/\bar\gamma),
\end{equation}
where $\mathsf E_m(z)=\int_1^\infty e^{-zt}t^{-m}\,dt$
\cite{dlmfexp} and $M_0=1$. The moment expression follows by
substituting $t=1+g$ in the integral of
$(1+g)^{-m}e^{-g/\bar\gamma}/\bar\gamma$ over $g\ge0$.

Expanding $(1-t)^{K-1}$ in the integral defining $I_x(K,K)$
gives powers $x^{K+j}$ with coefficients
$(-1)^j\binom{K-1}{j}/[(K+j)B(K,K)]$. Substituting
$x=c+d/(1+\gamma)$ and expanding each power expresses the
fading average in terms of $M_m$:
\begin{align}
\bar P_b
&=\frac{1}{B(K,K)}
\sum_{j=0}^{K-1}\frac{(-1)^j\binom{K-1}{j}}{K+j}
\nonumber\\
&\quad\times
\sum_{m=0}^{K+j}\binom{K+j}{m}c^{K+j-m}d^m M_m.
\label{eq:ray}
\end{align}
Powers with exponent zero are taken as one, including at the
endpoints. For $r<1$, bounded convergence gives the limiting
BEP $I_{(1-r)/2}(K,K)$ as $\bar\gamma\to\infty$. At $r=1$,
the high-SNR behavior is
\begin{align}
\bar P_b &\sim
\begin{cases}
(\log\bar\gamma)/(2\bar\gamma), & K=1,\\
C_K/\bar\gamma, & K>1,
\end{cases}\\
C_K&=\frac{B_{1/2}(K-1,K)}{2B(K,K)}-\frac12,
\qquad K>1,
\end{align}
where $B_x(a,b)$ denotes the unregularized incomplete beta
function.

For $K=1$, $\bar P_b=M_1/2$ at $r=1$, and the result follows
from the small-argument expansion of $\mathsf E_1$ \cite{dlmfe1}.
For $K>1$, \eqref{eq:asym} makes the conditional BEP integrable
over $g\ge0$. Dominated convergence then gives
\begin{equation}
\lim_{\bar\gamma\to\infty}\bar\gamma\bar P_b
=\int_0^\infty I_{1/[2(1+g)]}(K,K)\,dg=C_K.
\end{equation}
Substituting $x=1/[2(1+g)]$ and integrating by parts yields
\begin{align*}
C_K
&=\frac12\int_0^{1/2}\frac{I_x(K,K)}{x^2}\,dx\\
&=-I_{1/2}(K,K)\\
&\quad+\frac{1}{2B(K,K)}
\int_0^{1/2}x^{K-2}(1-x)^{K-1}\,dx.
\end{align*}
The boundary term at zero vanishes because
$I_x(K,K)=O(x^K)$ and $K>1$. Symmetry gives
$I_{1/2}(K,K)=1/2$, and the remaining integral is
$B_{1/2}(K-1,K)$.

At $r=1$, the conditional diversity order is $K$, while the
Rayleigh-averaged diversity order is one for every $K\ge1$.
Increasing $K$ adds pair observations that share the same
fading coefficient. For $K=1$, the logarithmic factor changes
the asymptotic decay rate while preserving diversity order one.

\section{Numerical Results}\label{sec:results}
\begin{figure}[t]
\centering
\includegraphics[width=.86\columnwidth]{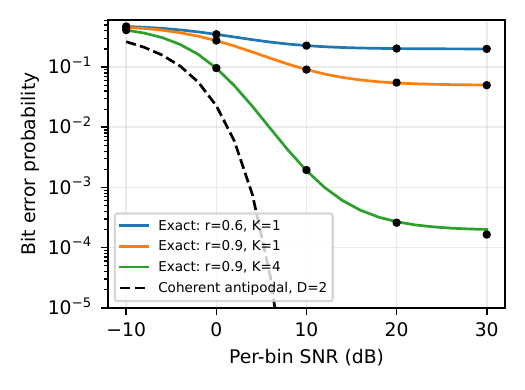}
\caption{Conditional BEP for the scalar channel. Curves show analytical results, and markers show Monte Carlo estimates.}
\label{fig:scalar}
\end{figure}
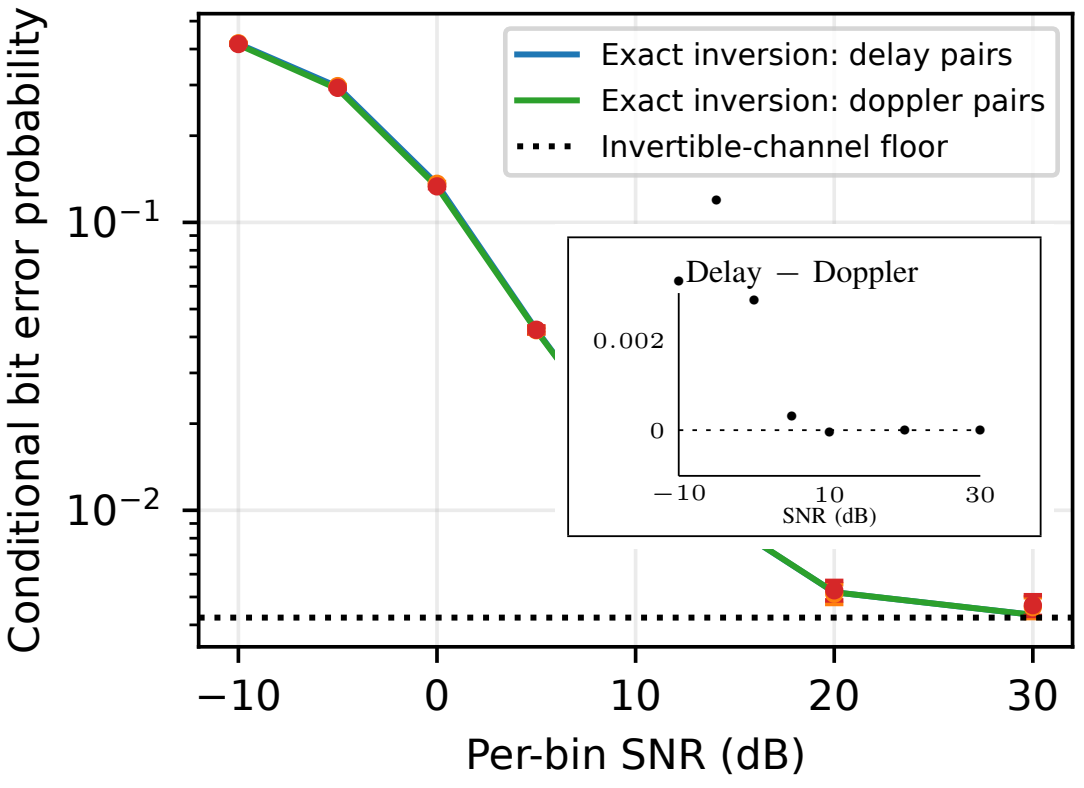
\begin{figure}[t]
\centering
\input{figures/pairing_comparison.tex}
\caption{Conditional BEP for delay and Doppler pairing. Curves show
analytical results, and markers show Monte Carlo estimates.
\textcolor{black}{Inset: exact delay-minus-Doppler BEP difference.}}
\label{fig:matrix}
\end{figure}
\begin{figure}[t]
\centering
\includegraphics[width=.86\columnwidth]{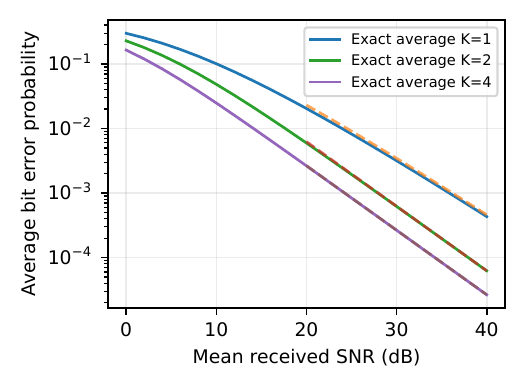}
\caption{Rayleigh-averaged BEP for $r=1$. Solid curves show exact results, and dashed curves show high-SNR asymptotes.}
\label{fig:ray}
\end{figure}

In this section, equiprobable binary states are used. So, each decision carries one bit across $L$ independently generated frames. The scalar tests use $K=JL$ pair observations, while the matrix tests use the full covariances in \eqref{eq:sigma}.

Fig.~\ref{fig:scalar} shows close agreement between the exact
scalar-channel BEP and the Monte Carlo estimates. For $K=1$,
increasing the correlation magnitude from $r=0.6$ to $r=0.9$
reduces the high-SNR error floor from $0.20$ to $0.05$. At $r=0.9$,
increasing $K$ from $1$ to $4$ lowers the floor to
$1.94\times10^{-4}$. At fixed transmit SNR, this improvement
comes with a fourfold increase in useful-frame energy per bit.
\textcolor{black}{The scalar markers use $2\times10^5$ trials each. The coherent antipodal reference matches the $K=1$ payload and useful-frame energy, with $E_b=2P$ and BEP $Q(\sqrt{4\gamma})$, where $Q(\cdot)$ is the standard Gaussian upper-tail function.}

Fig.~\ref{fig:matrix} compares the exact conditional BEP with
Monte Carlo estimates for a four-path fractional-Doppler OTFS
channel with $M=N=4$, $r=0.6$, $L=1$, and a three-sample CP.
\textcolor{black}{The path delays are $(0,1,2,3)$ samples, the normalized
Dopplers are $(0,0.35,-0.60,1.20)$, and the gains are proportional to
$(1,0.45e^{j0.7},0.30e^{-j0.8},0.20e^{j1.4})$, scaled so that
$\|\HH\|_F^2/N_{\mathrm{DD}}=1$.}
\textcolor{black}{The delay codebook pairs delay indices $(0,1)$ and $(2,3)$ at each Doppler index, and the Doppler codebook pairs the same index sets along Doppler at each delay index. Each carries one bit using eight pairs and energy $16P$ per useful frame. The markers use $6\times10^4$ trials per state and differ from inversion by at most $1.90$ estimated standard errors. The inversion uses absolute and relative quadrature tolerances of $2\times10^{-9}$ and $2\times10^{-8}$. The exact delay and Doppler BEPs are $0.13586291$ and $0.13296033$ at $0$~dB, and $0.01446809$ and $0.01451672$ at $10$~dB. The exact ordering reverses with a gap of $4.86\times10^{-5}$ at $10$~dB. Both approach the common floor of $0.00424$.}
\textcolor{black}{For delay pairing at $10$~dB, a separate sampled-waveform
simulation with explicit CP insertion and removal gave $0.01421$ from
$4\times10^4$ trials per state, compared with $0.01447$ from inversion.}
To assess whether the pairing advantage persists across channels,
both codebooks are evaluated on the same 64 channel realizations
with the same path delays and Dopplers as in Fig.~\ref{fig:matrix}.
\textcolor{black}{Independent zero-mean proper Gaussian path gains had
variances proportional to $(1,0.45^2,0.30^2,0.20^2)$, with unit total
expected power and no per-realization normalization. The NumPy random generator is initialized with seed 20261104 in the accompanying reproduction script.}
At $10$~dB, the mean conditional BEPs are $0.02552$ for delay
pairing and $0.02545$ for Doppler pairing. The mean paired
difference (delay minus Doppler) is $6.74\times10^{-5}$, with a
95\% paired $t$ confidence interval of $[-3.50,4.85]\times10^{-4}$.
The interval includes zero, so the tested ensemble shows no clear
average BEP advantage for either pairing.

Fig.~\ref{fig:ray} shows the Rayleigh-averaged BEP at $r=1$ when
all $K$ pair observations share a common fading coefficient.
Increasing $K$ lowers the BEP at a given average SNR. For $K>1$,
the BEP decays as $C_K/\bar\gamma$, whereas for $K=1$ it decays
as $(\log\bar\gamma)/(2\bar\gamma)$. Thus, all cases have diversity
order one. The logarithmic factor slows the approach to the
asymptotic slope for $K=1$.

\section{Conclusion}
This letter investigated the error performance of ND-OTFS with paired covariance signaling. It derived
the ML detector for a finite covariance alphabet and obtained the exact conditional BEP for binary signaling over a known effective DD channel. For scalar channels, the analysis yielded an incomplete-beta expression for the conditional BEP and an exact finite-sum expression for its Rayleigh average. The high-SNR analysis showed
that partial correlation produced a conditional error floor, whereas full correlation yielded a conditional BEP that decayed with an SNR exponent equal to the number of pair observations. \textcolor{black}{At $r=1$, the diversity order under common scalar Rayleigh fading was one.} Monte Carlo and sampled-waveform simulations corroborated the analytical results. \textcolor{black}{For one fixed fractional-Doppler channel, the exact pairing order reversed with SNR. The tested channel ensemble showed no clear average BEP advantage at equal payload and useful-frame energy.}

\makeatletter
\newif\ifrevisionbib
\let\originalbibitem\bibitem
\renewcommand{\bibitem}[1]{%
  \ifrevisionbib\egroup\revisionbibfalse\fi
  \originalbibitem{#1}%
  \def\entrykey{#1}%
  \def\dbnkey{dbn}\def\diversitykey{dbndiversity}\def\rcpkey{raviteja_rcp}%
  \ifx\entrykey\dbnkey\revisionbibtrue\fi
  \ifx\entrykey\diversitykey\revisionbibtrue\fi
  \ifx\entrykey\rcpkey\revisionbibtrue\fi
  \ifrevisionbib\bgroup\color{black}\fi}
\let\originalendbibliography\endthebibliography
\renewcommand{\endthebibliography}{%
  \ifrevisionbib\egroup\revisionbibfalse\fi
  \originalendbibliography}
\makeatother
\bibliographystyle{IEEEtran}
\bibliography{references}
\end{document}

%% file: figures/transceiver.tex
\setlength{\unitlength}{1pt}
\begin{picture}(248,103)
\put(0,83){\scriptsize $q$}
\put(8,85){\vector(1,0){10}}
\put(18,70){\framebox(61,30){\shortstack{\scriptsize \textcolor{black}{Gaussian}\\\scriptsize \textcolor{black}{mapper $\mathbf K_q$}}}}
\put(79,85){\vector(1,0){24}}
\put(84,90){\scriptsize $\mathbf x_\ell$}
\put(103,70){\framebox(61,30){\shortstack{\scriptsize OTFS $\mathbf A$\\\scriptsize + CP insertion}}}
\put(164,85){\vector(1,0){23}}
\put(167,90){\scriptsize $\mathbf s_\ell^{\rm CP}$}
\put(187,70){\framebox(61,30){\shortstack{\scriptsize Delay--Doppler\\\scriptsize channel + AWGN}}}
\put(218,70){\vector(0,-1){31}}
\put(187,9){\framebox(61,30){\shortstack{\scriptsize CP removal\\\scriptsize + OTFS $\mathbf A^H$}}}
\put(187,24){\vector(-1,0){23}}
\put(170,29){\scriptsize $\mathbf y_\ell$}
\put(103,9){\framebox(61,30){\shortstack{\scriptsize Collect $L$\\\scriptsize DD frames}}}
\put(103,24){\vector(-1,0){24}}
\put(18,9){\framebox(61,30){\shortstack{\scriptsize ML detector\\\scriptsize \textcolor{black}{$\{\mathbf{\Sigma}_i\}_{i=0}^{Q-1}$}}}}
\put(18,24){\vector(-1,0){10}}
\put(0,29){\scriptsize $\hat q$}
\end{picture}

%% file: figures/pairing_comparison.tex
\begingroup
\setlength{\unitlength}{1pt}
\resizebox{.86\columnwidth}{!}{%
\begin{picture}(252,187.2)
\put(0,0){\includegraphics{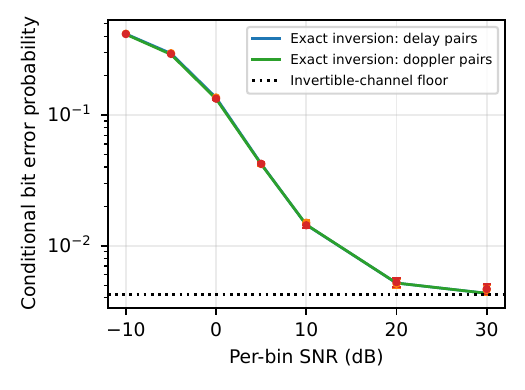}}
\put(130,64){\colorbox{white}{\textcolor{black}{\framebox(103,65){}}}}
{\color{black}
\put(184,121){\makebox(0,0){\scriptsize Delay $-$ Doppler}}
\put(157,77){\line(1,0){66}}
\put(157,77){\line(0,1){40}}
\put(154,87){\makebox(0,0)[r]{\tiny $0$}}
\put(154,106.6){\makebox(0,0)[r]{\tiny $0.002$}}
\multiput(157,87)(3,0){22}{\line(1,0){1}}
\put(157,73){\makebox(0,0){\tiny $-10$}}
\put(190,73){\makebox(0,0){\tiny $10$}}
\put(223,73){\makebox(0,0){\tiny $30$}}
\put(190,68){\makebox(0,0){\tiny SNR (dB)}}
\put(157.000,119.616){\circle*{2.2}}
\put(165.250,137.460){\circle*{2.2}}
\put(173.500,115.445){\circle*{2.2}}
\put(181.750,90.065){\circle*{2.2}}
\put(190.000,86.523){\circle*{2.2}}
\put(206.500,86.981){\circle*{2.2}}
\put(223.000,87.000){\circle*{2.2}}
}
\end{picture}}
\endgroup